\documentclass{article}
\title{Understanding Deutsch's Probability in a Deterministic Multiverse}
\author{Hilary Greaves*}
\date{}

\newcommand{\naive}{na\"{\i}ve}

\begin{document}

\maketitle

\begin{figure*}
Department of Philosophy, Rutgers University, 26 Nichol Avenue, New Brunswick, NJ 08901-2882, USA. \textit{email:} hgreaves@rci.rutgers.edu.

(July 11, 2003; revised April 29, 2004)
\end{figure*}

\begin{abstract}

Difficulties over probability have often been considered fatal to the Everett interpretation of quantum mechanics. Here I argue that the Everettian can have everything she needs from `probability' without recourse to indeterminism, ignorance, primitive identity over time or subjective uncertainty: all she needs is a particular \emph{rationality principle}.

The decision-theoretic approach recently developed by Deutsch and Wallace claims to provide just such a principle. But, according to Wallace, decision theory is itself applicable only if the correct attitude to a future Everettian measurement outcome is subjective uncertainty. I argue that subjective uncertainty is not to be had, but I offer an alternative interpretation that enables the Everettian to live without uncertainty: we can justify Everettian decision theory on the basis that an Everettian should \emph{care about} all her future branches. The probabilities appearing in the decision-theoretic representation theorem can then be interpreted as the degrees to which the rational agent cares about each future branch. This reinterpretation, however, reduces the intuitive plausibility of one of the Deutsch-Wallace axioms (Measurement Neutrality).

\end{abstract}


\begin{section}{Introduction and overview}
\label{intro}

The quantum measurement problem is very simple: under the assumption that our measurement apparatus is in principle describable by quantum mechanics, the collapse postulate contradicts the unitary dynamics. The Everett solution of the measurement problem is equally simple, on the face of it: discard the collapse postulate. This is the `conservative' solution, in the sense that it is the only solution that retains the existing physics (quantum state and unitary evolution) without amendment or addition. Such conservatism is surely desirable: quite apart from the issue of elegance, the existing quantum dynamics has been overwhelmingly empirically successful, and the minute we change the dynamical equations, we have no guarantee that results derived using the old equations will still hold. Further, such dynamical supplements invariably spoil the relativistic covariance of the theory; if we can successfully interpret the dynamics as it is, covariance is guaranteed. We thus have a powerful prima facie motivation for preferring the Everett interpretation. (Saunders (1997).)

But the Everett interpretation faces its own demons. How are we to understand the worldview it gives us?  In particular: (i) if it requires us to hold that the world `splits', what can determine that the `splitting' will occur along the lines prescribed by one basis in Hilbert space rather than another (the preferred basis problem)? (ii) How can we make sense of personal identity over time in an Everettian picture? (iii) Isn't this all unacceptably ontologically extravagant? (iv) Isn't it in any case just plain \emph{unbelievable}? (v) What about probability?

This paper is a discussion of (v), but first, cards must be laid on the table regarding (i)---(iv). In response to the preferred basis problem, in particular, there is a growing tradition of attempting to flesh out the basic Everettian proposal (Everett (1957)) by appeal to \emph{decoherence} (e.g. Gell-Mann and Hartle (1990), Saunders (1993, 1995), Wallace (2002a, 2003a), Zurek (1991, 1993, 1998)). According to this tradition, we should indeed consider that the world `splits' when a quantum measurement is performed: after the measurement, each possible outcome will be realized in some branch of the ever-proliferating universe (`multiverse', if we prefer), and (Wallace (2002a)) each branch deserves to be called a separate `world'. But, crucially, we are to understand this splitting as an effective and approximately defined process emergent from the quantum dynamics: a `measurement' is just a physical process in which a microscopic superposition is magnified (under the unitary dynamics) up to a macroscopic scale, and a `split' is just the advent of a suitable large-scale superposition. It is decoherence, rather than any \emph{primitive} `global bifurcation of spacetime' (\emph{pace} Earman 1986, p.224), that prevents parallel branches from subsequently interfering with one another. No heavy metaphysics, and no multiplication of mass-energy (\emph{pace} Healey (1984)), is to be introduced.\footnote{The decoherence approach is to be contrasted with a `further fact' many-worlds interpretation. According to the latter, the dynamics is explicitly supplemented by interpretive postulates that specify by \emph{fiat} which parts of the wavefunction correspond to distinct worlds; two proposals that agreed on the universal state but disagreed on the delineation of worlds would then be individually coherent, but mutually incompatible, proposals. It is in fact difficult to find \emph{proponents} of the Everett interpretation who advocate a further fact interpretation --- Deutsch (1985) is an exception, although Deutsch himself no longer holds the view expressed in that paper. A further fact view is often attributed to DeWitt (1970), but I can find nothing in his writings to generate any tension with (instead) a decoherence-based understanding of `splitting'. However this may be, the further fact MWI is pedagogically useful, not least because it is the understanding of Everett that seems to be envisaged by several \emph{critics} of `the' MWI (e.g. Kent (1990), Albert and Loewer (1988), Lewis (2004)).

Hereinafter, I will sometimes refer to the interpretation in question as `Everett'. This is simply an abbreviation; I will always be referring to the the decoherence-based picture sketched above rather than to the person, and at no point do I intend to make any historical claim regarding Everett's original (1957) ideas.}

 The spirit of the decoherence approach to the preferred basis problem, clearly, is that \emph{everything supervenes on the existing physics}. As such, it goes naturally hand-in-hand with particular responses to the next two of the challenges listed above. We are to adopt a reductionist approach to personal identity over time and a fairly strong supervenience of the mental on the physical: once the physics has been specified, there is no \emph{further} matter of which post-measurement observer will be \emph{me}, and we are simply to choose a best description (here we part company with `many-minds' approaches, e.g. Albert and Loewer (1988)).

 The next two challenges are relatively easy to dispose of. To the charge of ontological extravagance it can be countered that, firstly, such `extravagance' is preferable to the \emph{theoretical} extravagance and inelegance required to \emph{eliminate} other branches when our best formalism predicts their existence; secondly, this ontological extravagance is anyway not too damaging when the extra entities are of the same \emph{kind} as those already admitted to existence. Complaints that the Everettian picture is simply unbelievable may be well taken, but as far as philosophy goes, we can reply only by borrowing David Lewis's memorable remark: ``I cannot refute an incredulous stare."

 So far so good, perhaps. But even if the Everettian can get this far (which remains an open question --- see, e.g., Zurek (2003) for a recent review of the state of play vis-\`{a}-vis the preferred basis problem), she will still face a problem over the issue of \emph{probability} --- what story is to be told about the relationship between the probabilistic statements of instrumentalist QM, and the Everettian worldview? This is the topic of the remainder of this paper; from here on, I make the working assumption that the decoherence-based approach is, at least \emph{otherwise}, sound.

 We can begin by distinguishing two aspects of the probability problem. The first is the \emph{incoherence problem}. Everettian quantum mechanics appears to be a straightforwardly deterministic theory --- there is, in Everett, no necessary ignorance of initial conditions (in contrast to the de Broglie-Bohm pilot wave theory --- see, e.g., Bohm (1952)), and no irreducibly stochastic collapse (in contrast to state-reduction theories --- see, e.g., GRW (1986), Pearle (1989)). Instead, we can be certain that each possible outcome of a quantum measurement is realized in some post-measurement branch. In that case, can it even \emph{make sense} to talk of probability?

 The second is the \emph{quantitative problem}, which remains even if the incoherence problem can be solved: can Everett recover probabilities that \emph{numerically agree} with those of the Born rule? The quantitative problem is perhaps unproblematic in the case of equally weighted superpositions, but has been considered fatal in the case of unequally weighted superpositions. The reason for the latter is that when a superposition is unequally-weighted, the instrumentalist QM algorithm would assign \emph{unequal} probabilities to the various outcomes. In contrast, it seems that all Everett can say is that each outcome occurs in exactly one branch\footnote{Actually, it is overly {\naive} to suppose that the Everettian can strictly even say this: the `number of branches', on a decoherence-based approach to splitting, will not be well-defined. This will be important later --- see section \ref{failureofegalitarianism}.}, which, we might think, will yield \emph{equal} probabilities if any at all.

 Now, one might think that any would-be interpretation of quantum mechanics is utterly constrained to recover probability properly so-called (whatever that may mean), simply in order to deserve the name. In that case the challenge to Everett would seem to be a straightforward ultimatum: probability or die. But we should be more careful than this. Any theory that recovers the virtues of QM (or equivalent virtues) while avoiding its vices is acceptable; if it deserves the title `replacement for' rather than `interpretation of' quantum mechanics, so be it.

 According to the view I will advocate (\textbf{section \ref{theproblem}}), what is required, in order to render the Everett interpretation acceptable, is a particular \emph{rationality principle}: the Everettian must be advised to care about her future branches in proportion to their relative quantum amplitude-squared measures. Such a principle will do two things for the Everettian: it will tell her how to act in the face of the otherwise daunting picture of branching personal identity that Everett gives us, and, on a plausible (albeit not self-evident) account of empirical confirmation, it will allow her to regard quantum mechanics (understood in her Everettian way) as an empirically confirmed theory. \emph{If} we think that probability is \emph{defined} by its role in prescribing rational action, then this amounts to recovering `probability'; under other construals of `probability' it may not; but \emph{that} question is merely terminological.

 In that case the decision-theoretic approach of Deutsch (1999) and Wallace (2002b/2003b) is precisely called for; in \textbf{section \ref{decisiontheory}} I give a brief overview of the decision-theoretic programme and its application to the Everett interpretation. Deutsch, and Wallace following him, invokes decision theory in an attempt to prove that a rational Everettian should act `as if' the Born probability rule were true. \emph{If} this proof is sound, and \emph{if} (in addition) my arguments of section \ref{theproblem} are correct, then the problem of probability in the Everett interpretation has been solved.

 Sections \ref{applicability} and \ref{mn} offer critical discussion of the decision-theoretic approach. \textbf{Section \ref{applicability}} addresses the question of whether decision theory is even \emph{applicable} to the Everettian branching scenario. We might think not: according to Wallace, for instance, the intuitive and conceptual acceptability of the proof rests upon the premise that a rational Everettian facing an imminent measurement, despite knowing all the relevant facts concerning her situation (that she will split, etc.), should regard her situation as one of \emph{subjective uncertainty} (SU). In section \ref{su} I argue that this premise is false. I claim, instead, that (given our reductionist approach to personal identity over time) the rational Everettian should feel that she knows exactly what is going to happen, \emph{including} what is going to happen \emph{to her}. In that case, one might think that decision theory is inapplicable to Everettian branching --- decision theory is, after all, a theory designed for decision-making \emph{under uncertainty} --- and thus that Deutsch's approach is misguided; one might further think that, `proved' or not, the rationality principle itself just doesn't make sense in an Everettian context. I argue against each of these suspicions. I claim (section \ref{cf}) that we can understand the reasonableness of the decision-theoretic rationality constraints on the basis of the fact that the rational Everettian agent should \emph{care about} all of her futures; the `probabilities' appearing in Deutsch's decision theory are then understood as measures of the degree to which the present person-stage cares about each future branch. In \emph{this} respect, the decision-theoretic program is not damaged by the abandonment of the `subjective uncertainty' claim.

 In \textbf{section \ref{mn}} I turn to a more contentious aspect of Deutsch's proof. Deutsch must assume, in addition to decision theory, an axiom that Wallace christens `Measurement Neutrality' (MN): that the rational agent is indifferent between any two quantum `bets' (or `acts' or `games'; see section \ref{quantumrepresentationtheorem}) that involve measuring the same observable $\hat{X}$ on the same state $\vert \psi \rangle$, and assigning the same reward as each other to each observed eigenvalue, but disagree on \emph{how} $\hat{X}$ is to be measured on $\vert \psi \rangle$. That this (originally tacit) assumption is crucial to Deutsch's proof, and that it is more contentious than a first glance suggests, was pointed out by Wallace (2002b/2003b). But Wallace suggests that the prospects for giving MN a strong intuitive justification are good. I will argue that while this \emph{may} be true under a sufficiently strong SU philosophy, the prospects look bleak if, as I urge, we abandon SU.

 This is not to say that the probability problem is insoluble, and hence the Everett interpretation sunk. Neither is that my position: in section \ref{spectrum} I argue that in fact, we should accept as something of a \emph{primitive} that rational action for an Everettian involves acting `as if' the Born rule is true. This move, however, leaves the decision-theoretic program playing a somewhat different role from that originally intended by Deutsch: rather than persuading a would-be dissenter of the truth of the Born rule, the decision-theoretic program serves as an illustration of the internal coherence of such primitive acceptance.

 \textbf{Section \ref{conclusions}} summarizes the paper's conclusions. The demand for `probability' in any more exalted sense than that of decision theory is (with one important proviso) misguided. There is no conceptual obstacle to applying decision theory to Everettian branching: the very same reasons that urge a non-branching agent to care about her unique future urge an Everettian to care about all her futures, and this is enough. We need not and should not invoke uncertainty. But without uncertainty, Deutsch and Wallace's assumption of `Measurement Neutrality' is unjustifiable; the Everettian is thus driven to a more primitive acceptance of the Born rule. This, however, is not much of an objection to the Everett interpretation.

 \end{section}

 \begin{section}{The problem of probability} \label{theproblem}

 Discussions of the problem of probability in Everett usually begin from the tacit assumption that if Everett cannot sustain the probabilistic statements of `standard' QM, Everett must be rejected. The issues will be clearer if we start from one step further back: what, precisely, is the rationale for such conditional rejection?

 It is instructive to consider which features of Everett would \emph{survive} the loss of a probability interpretation. We are given a picture of an ever-branching universe, initially factorizable joint states of composite observer-object-environment systems evolving into entangled states. In those entangled states, each superposed component of the state of the joint system, in the decoherence-preferred basis, describes something that looks like (and indeed is) an approximately classical world; after a measurement, each branch describes a post-measurement state corresponding to one particular definite measurement outcome. This much will survive. So also will the consistency of the Everettian story with experiment. We have not observed \emph{probabilities}, we have observed relative frequencies, and these are reproduced by Everett: the Everett interpretation predicts that there will exist a branch with precisely the sequence of outcomes, and \emph{a fortiori} the relative frequencies, that we have in fact observed. (Everett also, of course, predictes that there will be branches collectively representing every other `possible' set of relative frequencies. I do not deny that herein lies \emph{a} problem: see section \ref{confirmation}. My point here is just that the problem is not inconsistency with experiment.)

 What of the quantum-mechanical amplitudes? Each component of the decomposition is associated with a complex number $c_i$ such that $0 \leq \vert c_i \vert \leq 1$, but, we are supposing, the $\vert c_i \vert ^2$ are not interpretable in terms of probability. The $c_i$ \emph{do} have physical (non-probabilistic) consequences in cases of \emph{interference}, but we neglect the possibility of interference between macroscopic branches.\footnote{One might try to justify this move by claiming (with support from the theory of decoherence) that macroscopic inter-branch interference is highly `improbable' relative to any plausible ignorance probability distribution over the set of initial states of the universe; in that way we would have `improbability' in the sense required for current purposes without interpreting the \emph{amplitudes} in terms of probability. But actually, it is questionable whether this move is valid - it may be that interference is \emph{common} between macroscopic branches of extremely low amplitude (thanks to David Wallace for pointing this out). If that is the case, we have a further need for something like probability, to justify neglecting branches of low amplitude, and hence to underwrite a decoherence-based solution to the preferred basis problem.} In that case, \emph{all} that can be said of the future relative to the pre-measurement observer's world is the following: all and only those outcomes that correspond to an eigenstate that has nonzero amplitude in the pre-measurement state of the microsystem will be realized, and thence observed, in some post-measurement branch. It is statements like \emph{these} that are the core empirical predictions of collapse-free, probability-free quantum mechanics; is this good enough? I will argue that it is not, but we need to be clear as to why (exactly!) such an impoverished interpretation of no-collapse quantum mechanics is inadequate.

 \begin{subsection}{What should I do?} \label{howtoact}

 On a purely pragmatic level, in the first instance, the recent convert to the Everett interpretation is faced with a bewildering dilemma - how is he rationally to act, believing that he has more than one future self? Dilemmas such as this have been considered in a (mostly) hypothetical tone in the philosophical literature (e.g. Parfit (1984)). One response (e.g., Wilkes (1988)) suggests that the scenario is technologically impossible, and that therefore we need not have an answer. However this may be in the classical case, such a comfortable reply is not available to the Everettian --- far from being \emph{impossible}, he believes, fission is \emph{inevitable}!

 This is an unavoidable philosophical problem for an Everettian. We are accustomed to believing that future quantum measurement outcomes have (whatever exactly this may mean) objective probabilities of actually occurring, and that in deciding on courses of action, we should (for whatever reason) weight our expectations according to those probabilities. (A toy situation in which this issue is urgent is discussed in section \ref{cf}.) If the amplitude-squared measure in the Everett interpretation does not bear a similar link to rational action, therefore, the `rational' foundation that the recent convert is accustomed to using to guide his behavior is kicked away, and a replacement guide, one that delivers the same advice \emph{or otherwise}, is urgently required --- inaction is impossible.

 \end{subsection}

 \begin{subsection}{Inadequate justification for believing Everett} \label{confirmation}

 \begin{quote} If a theory fails to be empirically coherent, it might be true, but if true, then one could never have empirical reason for accepting it as true. (Barrett (1999), pp.116-7) \end{quote}

 The issue of confirmation in Everettian quantum mechanics\footnote{I am not referring to the issue of the empirical confirmation of Everett \emph{as opposed to rival interpretations of quantum mechanics}. There are plenty of reasons for preferring the Everett interpretation over rivals on theoretical rather than empirical grounds (aesthetic appeal, economy on assumptions, prospects for relativistic covariance, and so on). Further, we can in fact find differences in empirical predictions between rival `interpretations', if we look hard enough (for instance, the Everett interpretation permits macroscopic interference phenomena that would be forbidden by the GRW theory). Rather, the worry is this: it may be that if, attracted (say) by such theoretical virtues, we decide to understand quantum mechanics along Everettian lines, suddenly we lose the empirical reason we had for believing \emph{quantum mechanics} in the first place.} is a tricky one. On the one hand, quantum mechanics, when understood along Everettian lines, makes very strong predictions. It is, after all, a deterministic theory. Given the state of the universe at any one time, an Everettian Laplace's Demon could tell us, in principle, exactly what will happen: which branches will exist and just what each will be like. From a God's-eye view, then, Everett is eminently falsifiable. The problem is that Everettian QM \emph{also} predicts that any individual observer at any given time (or community of interacting observers) will have a severely limited access to the multiverse. Each can interact with, and hence can observe, only the inhabitants of his (their) own branch; each can have interacted with, and hence can have memories or records of, only inhabitants of his (their) own branch's history. As far as predictions for \emph{individual branches} go, therefore, Everett's predictions are, until and unless we have some interpretation of the branch amplitudes, not all that much more than `anything is possible'. How can we have empirical reason to believe a theory that just says that anything is possible?

 Here the similarity to indeterministic theories is striking. So let us consider how we treat confirmation in those theories: say, quantum mechanics under an instrumentalist collapse interpretation. Again, we can never regard such a theory as absolutely empirically refuted, as long as what we observe to happen is something that the theory deemed \emph{possible}. What we do in this indeterministic case, of course, is to adopt a \emph{probabilistic} criterion of confirmation: we regard the theory as empirically confirmed iff the results we observe are ones that (or, ones that belong to a suitable class that) the theory would have told us to expect \emph{with high probability}.

 This, then, is the second reason why the Everett interpretation requires a probability interpretation \emph{or something very similar}: without one, the Everett interpretation is in the predicament Barrett (quoted above) laments (although not for the reason Barrett gives in his book). Such a theory could be true --- it is neither incoherent nor inconsistent with the experimental evidence --- but if true we could have no justification for believing it.

 \end{subsection}

 \begin{subsection}{Linking and solving the two problems}
 \label{link}

 At first sight, it seems that the two genuine probability-related problems I have outlined above are rather different: on the one hand, we ask how we (rationally) should act, given the assumed truth of Everettian QM; on the other, we ask whether we (rationally) should believe Everettian QM in the first place.

 Against this, I propose that the following, if true, will suffice to kill \emph{both} of the above problems with one stone: \emph{the rational Everettian cares about her future successors in proportion to their relative amplitude-squared measures}.

 It is easy to see that this is relevant to the first problem, i.e. the problem of how rationally to act when facing the prospect of a branching future. If I know how to weight the interests of my successors one against the other, then I can calculate which course of action will best serve my overall interest. (These ideas are made precise by the machinery of decision theory, for which see section \ref{decisiontheory}.)

 On the face of it, it is far less easy to see that such a `caring measure' would have anything to do with empirical confirmation of theories. Nevertheless, I think a fairly strong case can be made for the following philosophical conjecture: \begin{quote} If the rational Everettian cares about her future successors in proportion to their relative amplitude-squared measures, then (given our actual experimental data) she should regard (Everettian) quantum mechanics as empirically confirmed. \end{quote}

 Why is this conjecture at all plausible? Well, consider (for instance) the Bayesian account of empirical confirmation. According to Bayesians, the rational agent updates his credence in a theory by Conditionalization:

 \begin{displaymath} p^A_{new}(T) = p_{old}(T/A) \equiv \frac{p_{old}(A/T)}{p_{old}(A)}p_{old}(T) \end{displaymath}

 where \begin{itemize} \item $T$ is the proposition that a certain theory (for instance, quantum mechanics) is true, \item $A$ is some proposition in which we might come to have degree of belief 1 (for instance, that the dial swung to point at `spin-up'), \item $p_{old}(\cdot)$ is the agent's credence function before obtaining information as to the truth of A, and \item $p^A_{new}(\cdot)$ is the agent's credence function after coming to have degree of belief 1 in A. \end{itemize}

 Why Conditionalize? Bayesians have a variety of responses to this challenge. One is a diachronic `Dutch Book' argument (e.g. Teller 1973): it can be shown that, if the agent updates his degrees of belief in \emph{any} nontrivial way on learning that A, and if his updating is other than by Conditionalization, then a `Dutch book' can be made against him. That is, one can devise a series of bets --- all of which the agent is committed to accepting (according to his announced degrees of belief at the time the bet is offered) --- such that the agent is guaranteed to make a net loss, no matter what the truth-values of the relevant propositions.

 How? Such a Dutch Book involves three bets. The first bet (which will be our primary concern here) is offered \emph{before} the truth of A is known: the agent is required to bet that A is true, using his current degree of belief in A as betting quotient. The truth or falsity of A is then ascertained, the agent updates his beliefs according to his chosen policy, and two further bets are offered using the agent's new betting quotients. Gains and losses from the three bets are summed to give the net result.

 My point is that \emph{a precisely analogous Dutch Book argument supports updating one's credence in a theory by Conditionalization in the deterministic branching case.} The fundamental assumption underpinning Dutch Book arguments is that betting quotients should equal degrees of belief. (Indeed, this is often regarded as \emph{definitive} of degrees of belief.) So, in the stochastic case, the agent's degree of belief in A serves as his betting quotient for A: it is this that commits him to accepting the first bet. But in the deterministic branching case, surely, the agent's \emph{caring measure} over his future branches plays exactly the same role in determining betting quotients. In the branching scenario, the agent \emph{knows} that he has two real future branches, on one of which A occurs (and which he cares about to degree $p$), and on the other of which $\neg$A occurs (and which he cares about to degree $(1-p)$). Accepting a bet (pre-measurement) on the truth of A will commit one successor to a certain gain, and another successor to a certain loss. There will be some betting quotient such that our agent is indifferent between accepting this bet and rejecting it: I claim that that betting quotient is $p$. (Again, we might well regard this as \emph{definitive} of the caring measure.) Therefore, if the agent, on observing A or $\neg$A, updates his credence in the theory nontrivially \emph{but other than by Conditionalization}, then one can offer a series of bets which he will accept but which guarantee him a net loss \emph{on every branch}. This is just as bad as being guaranteed a net loss \emph{no matter what the outcome}: what happens to successors on those future branches constitutes our agent's future. The branching agent, therefore, should Conditionalize for the same reason that the agent in a stochastic world should Conditionalize.

 \emph{If} we accept this argument, the problem for Everett is then in focus. Consider the degree to which, had I believed Everett before I heard the results of any quantum-mechanical experiments, it would not have been rational for me act as if I had a single successor and to regard a class of data of which the actual data are typical as highly probable. To that same degree, it is not rational to believe Everettian quantum mechanics.

 There is plenty more to be said in justification, and in criticism, of my conjecture; I do not pretend to have provided a conclusive defense of it. (In particular, it is worrying that the Dutch Book argument has nothing to say against an agent who insists that the observation of A or $\neg$A is \emph{irrelevant} to his belief in the theory, and hence that he will retain his old degrees of belief: $p^A_{new}(\cdot) = p_{old}(\cdot)$.) The relevance of the remainder of this paper to the second problem (i.e. to the issue of whether we should believe Everettian quantum mechanics) is conditional on the truth of this conjecture that a `caring measure' would play the same role as stochastic chance in theory confirmation. (One more supportive comment, though: we have no \emph{alternative} account of what it would take to empirically confirm a deterministic branching theory, and to say that there is \emph{no} way to confirm \emph{any} such theory would be --- I think --- suspicious.)

 It is crucial to note that if my above arguments as to what is and what is not problematic about the issue of probability in Everett are correct, the \emph{incoherence} problem is transformed, as follows. If it is really an indispensable criterion for applicability of the term `probability' that (say) one outcome must occur to the exclusion of the others, or that probability measures uncertainty, then perhaps there is no \emph{probability} in Everett, but this cannot in itself generate a \emph{problem}. In fact, I share Deutsch's view that the defining feature of `probability' is its role in prescribing rational action: in other words, if some given feature of reality plays the role that probability plays in rationality, then that feature of reality thereby deserves the name `probability'. I will therefore join standard practice in using the term `probability' within the decision-theoretic programme (below). But any reader sceptical of this view of `probability' should note that the disagreement is a purely verbal one: \emph{if} my above conjecture is correct, then the genuine problems in Everett arise from the need for a {rationality principle}, with or without title to the name `probability'.

 We \emph{would} have an `incoherence problem' if we could not defend such a rationality principle. For instance, it could turn out that there were \emph{no} coherent strategy for rational action in an Everettian context. In \emph{that} case, according to my sections \ref{howtoact} and \ref{confirmation}, Everett would be not only (a) useless in guiding our actions, but also (b) impossible to empirically confirm, due to `probability not making sense'. As Deutsch notes, this possibility cannot be ruled out of court: \begin{quote} It is not self-evident \ldots that rationality is possible at all, in the presence of quantum-mechanical processes --- or, for that matter, in the presence of electromagnetic or any other processes. (Deutsch (1999), p.3130) \end{quote} Further, we would have a problem if it should turn out that Everettian rationality, while possible, does not underwrite any representation theorem that uses a probability function (see section \ref{decisiontheory}). This, though, is rather different to the `incoherence problem' that much previous work on the topic has been concerned with (e.g. Albert and Loewer (1988), Loewer (1996), Saunders (1998), Wallace (2002b/2003b), Ismael (2003)).

 \end{subsection}

 \end{section}

 \begin{section}{The decision-theoretic programme} \label{decisiontheory}

 In the light of the above, the line pursued by Deutsch (1999) and Wallace (2002b/2003b), namely to address probability concerns in Everett by means of decision theory, is precisely the right approach --- the two genuine problems, that of justifying belief in Everett and the practical question of what to do if we do hold such belief, come down to the single issue of rational action in the face of an imminent measurement given the truth of Everett, and it is this issue that the decision-theoretic programme addresses.

 In section \ref{classicaldecisiontheory} I summarize the basic ideas behind decision theory as treated by Savage (1972). (See Joyce (1999) for a more extended exposition, and Fishburn (1981) for a thorough review of Savage's approach and alternatives.) In section \ref{quantumrepresentationtheorem} I give an informal sketch of Deutsch's proof that the `rational' Everettian agent is constrained to adopt the Born rule. (Here also, my account will be no more than a sketch; for the details, see Deutsch's original paper and/or Wallace's exegesis and extension.) I postpone criticism until sections \ref{applicability} and \ref{mn}.

 \begin{subsection}{Savage decision theory} \label{classicaldecisiontheory}

 Decision theory, as usually conceived, is a theory of rational action in the face of uncertainty: it provides a framework in which we can discuss an agent's choices of actions, provided that those choices satisfy certain intuitively reasonable constraints which we are prepared to regard as criteria of rationality. We begin by carving up the agent's decision problem by means of the following sets:

 \begin{itemize}

 \item A set $\mathcal{C}$ of \emph{Consequences}. Under the intended interpretation, this set is the locus of \emph{value to the agent}: some Consequences are better for the agent than others. Typical Consequences might be winning the London Marathon, receiving \$10, or contracting pneumonia.

 \item A set $\mathcal{S}$ of \emph{States} of the world. Under the intended interpretation, this set is the locus of the agent's \emph{uncertainty}: the agent is uncertain as to which State $S \in \mathcal{S}$ ``obtains". We then define the set $\mathcal{E}$ of \emph{Events} as the power set of $\mathcal{S}$. Typical examples of either States or Events (there is flexibility as to just how fine-grained the individual States are to be) might be a certain egg's being rotten, or its raining tomorrow.

 \end{itemize}

 The sets $\mathcal{C}$ and $\mathcal{S}$ are taken as basic primitives, and interpreted as indicated by the examples given. We form the set $\mathcal{C} ^\mathcal{S}$ of functions from $\mathcal{S}$ into $\mathcal{C}$, and define:

 \begin{itemize}

 \item A set $\mathcal{A} \subseteq \mathcal{C}^ \mathcal{S}$ of \emph{Acts}. These are to be interpreted as courses of action that an individual might make. The mathematical representation of and Act is by means of a function from States to Consequences because the agent is supposed to know which Consequence will result from each of the Acts between which he is choosing, given any State $S \in \mathcal{S}$. (If the agent does not have such complete knowledge of Consequences, this means only that we have carved up the decision problem incorrectly --- see Joyce (1999), pp.52-3.) Typical Acts might be crossing to road, playing a game of tennis, or purchasing a \$10 bet on Blue Murder at 16:1 odds.

 \item A weak preference relation $\succeq$ $\subseteq \mathcal{A} \times \mathcal{A}$, with the intended interpretation that for all Acts $A, B \in \mathcal{A}$, $A \succeq B$ iff the agent prefers Act $A$ to Act $B$ or is indifferent between $A$ and $B$. \end{itemize}

 Following Wallace (2002b), we add a set of \emph{Chance Setups}, enabling the theory to deal with the fact that the set of future States of the world is in general \emph{dependent} on the agent's choice of Act:

 \begin{quote} A set $\mathcal{M}$ of \emph{Chance Setups}, to be regarded as situations in which a number of possible events might occur, and where it is in general impossible to predict with certainty which will occur. Examples might be a rolled die (in which case there is uncertainty as to which face will be uppermost) or a section of road over a five-minute period (in which case there is uncertainty as to whether or not a bus will come along).

 For each $M \in \mathcal{M}$ we define a subset $\mathcal{S}_M \subseteq  \mathcal{S}$ of States which might occur as a consequence of the Chance Setup. (The Event space [$\mathcal{E}_M$] for $M$ is  \ldots the power set of $\mathcal{S}_M$.) (Wallace (2002b), p.5; capitalization added) \end{quote}

 Decision theory then \emph{defines} probability in terms of the agent's preferences, as follows. We consider an agent faced with a choice of \emph{constant Acts}, i.e. Acts that assign the same Consequence to every possible State $s \in \mathcal{S} _M$. We say that if, for two such Acts $c_1$, $c_2$, a rational agent weakly prefers $c_1$ to $c_2$ (i.e. $c_1 \succeq c_2$), then for arbitrary events $E_a$, $E_b$, that agent's probability for $E_a$ is higher than or equal to his probability for $E_b$ iff the agent weakly prefers the deal ($c_1$ if $E_a$ occurs and $c_2$ if $E_b$ occurs) to ($c_2$ if $E_a$ occurs and $c_1$ if $E_b$ occurs).

 So far, of course, we have at best only a \emph{qualitative} notion of probability, and in fact no guarantee that the `higher than' relation we have defined is an ordering. To gain such a guarantee and to proceed to a \emph{quantitative} definition, we require that the agent's preferences among Acts satisfy certain intuitively highly reasonable axioms: that the preferences be transitive (if an Act A is weakly preferred to B and B is weakly preferred to C, then A is weakly preferred to C), that they satisfy Dominance (if an Act A gives Consequences that are weakly preferred to those given by Act B on every possible State, then Act A is weakly preferred to Act B), and so forth.\footnote{The decision-theoretic axioms are far from uncontroversial within discussions of decision theory itself, and there is more than one possible set of axioms that one could use. I will refrain from further discussing the specifics: we can regard any serious candidate set of axioms as uncontentious for present purposes. Further, the choice is not crucial to the quantum project: Wallace (2002b) gives several variant proofs of the quantum representation theorem (see section \ref{quantumrepresentationtheorem}), from rival sets of axioms.}

 From these axioms, we prove a \emph{representation theorem}: that the agent's preferences among Acts determine (e.g. Savage (1972)) a unique quantitative `probability' measure $Pr_M: \mathcal{E}_M \rightarrow [0, 1]$ and (von Neumann and Morgenstern (1947); their axioms follow from Savage's) a value function $\mathcal{V}:\mathcal{C} \rightarrow \Re$, unique up to a multiplicative and an additive constant, such that the preferences are completely represented by the following rule.

 \paragraph{Expected Utility rule.} For all $A, B \in \mathcal{A}$, Act $A$ is preferred to Act $B$ iff $EU(A) > EU(B)$, where the \emph{expected utility} $EU:\mathcal{A} \rightarrow \Re$ is given by

 \begin{equation} \forall F \in \mathcal{A}, EU(F) = \sum_{x \in \mathcal{S}_M}Pr_M(x) \mathcal{V}(f(x)). \end{equation}

 Again: depending on one's philosophy of probability, one may or may not judge that the measure $Pr_M$ really deserves the name \emph{`probability'}, but for present purposes this should be regarded as a verbal matter.

 \end{subsection}

 \begin{subsection}{The Deutsch-Wallace quantum representation theorem} \label{quantumrepresentationtheorem}

 The decision-theoretic approach is then applied to Everettian quantum mechanics. We consider a pre-measurement observer who is faced with a choice of decision-theoretic \emph{Acts} (or `\emph{games}') involving quantum measurements. (`Measurement' is to be understood in the Everettian sense, i.e. as a process in which a microscopic system, perhaps in a state of superposition, becomes appropriately entangled with the state of (some macroscopic part of) the rest of the universe.) Each Act involves the agent agreeing to stake the fates of his future selves on the outcome of such a measurement, where the specification of the Act includes that of a particular \emph{state} $\vert \psi \rangle$ to be measured, \emph{observable} $\hat{X}$ to be measured on that state, and \emph{payoff function} $\mathcal{P}$ from observed eigenvalues to Consequences. We consider the agent's preferences among possible Acts, and require (in order that he be counted as rational) that those preferences satisfy the following constraints:

 (i) very close analogues of some set of conventional decision-theoretic axioms;

 (ii) \emph{Physicality} (P): the agent's preferences ultimately attach to physical scenarios rather than to mathematical objects, so that if two mathematical objects $\mathcal{G}_1 = \langle \vert \psi _1 \rangle , \hat{X} _1, \mathcal{P} _1 \rangle$, $\mathcal{G}_2 = \langle \vert \psi _2 \rangle , \hat{X} _2 , \mathcal{P} _2 \rangle$ represent the same physical scenario, the agent is indifferent between $\mathcal{G}_1$ and $\mathcal{G}_2$;

 (iii) \emph{Measurement neutrality} (MN): the agent is indifferent between two Acts that agree on $\vert \psi \rangle$, $\hat{X}$ and $\mathcal{P}$ but disagree on \emph{how} $\hat{X}$ is measured on $\vert \psi \rangle$.\footnote{Physicality and Measurement Neutrality are not made explicit in Deutsch's original paper, but, as Wallace points out, they are nontrivial (in the case of Physicality: not \emph{entirely} trivial) and essential to the proof. See section \ref{mn} for discussion of the acceptability of Measurement Neutrality.}

 It can then be proved (Deutsch (1999), Wallace (2002b/2003b)) that for an agent to be `rational' in the sense of these axioms, \emph{he is constrained to adopt the amplitude-squared measure as his probability function}.

 In outline, Deutsch's original proof proceeds as follows. (See Wallace (2002b), pp.38-41, or (2003b), pp.14-17, for a more detailed reconstruction, with those of the above assumptions that were tacit in Deutsch's work made explicit.) First, we consider an equally-weighted two-component superposition $\vert \psi \rangle = \frac{1}{\sqrt{2}} ( \vert x_1 \rangle + \vert x_2 \rangle )$. We consider a game (Act) in which an observable $\hat{X}$ whose spectrum includes $x_1$ and $x_2$ is to be measured on $\vert \psi \rangle$, and in which the successor who observes the result $x_1$ ($x_2$ resp.) will be awarded a Consequence $c_1$ ($c_2$ resp.); we ask how much the agent values this game. (`Value' is also to be understood in the decision-theoretic sense defined above --- recall that the point of a value function is to plug into the expected utility rule, and thus to encode the agent's preferences between Acts. Heuristically, as Deutsch notes, we can understand the value of a quantum game to be the greatest amount of money our rational agent would be willing to pay for the privilege of playing the game.) Using considerations of symmetry (which do not, however, invoke any assumptions \emph{over and above} those listed above), it is proved that the value of this game must equal the average of the values of the two Consequences (Deutsch's equation (2.10)/Wallace's Stage 1): \begin{equation} \mathcal{V}(\frac{1}{\sqrt{2}} ( \vert x_1 \rangle + \vert x_2 \rangle )) = \frac{1}{2}(\mathcal{V}(c_1) + \mathcal{V}(c_2)). \end{equation} Therefore, the agent must assign decision-theoretic probability $\frac{1}{2}$ to each measurement outcome. It is then straightforward to extend this result to an $n$-component equal-amplitude superposition for arbitrary $n$ (Deutsch's equation (3.1)/Wallace's Stage 2): \begin{equation} \mathcal{V}(\frac{1}{\sqrt{n}} ( \vert x_1 \rangle + \ldots + \vert x_n \rangle )) = \frac{1}{n} (\mathcal{V}(c_1) + \ldots + \mathcal{V}(c_n)). \end{equation}

 Second, we consider \emph{unequally}-weighted superpositions. Here Deutsch's strategy is to show that, on the Everettian understanding of `measurement', a measurement of an observable on an unequally-weighted superposition can be realized by a physical process that \emph{also} meets the definition of a measurement of another observable, $\hat{Y}$ say, on a system that is in an equally-weighted superposition of eigenstates of $\hat{Y}$. It then follows, from the fact that all eigenvalues of $\hat{Y}$ are to be considered equally probable (are to be assigned equal importance in the agent's calculation of the decision-theoretic value of the game), that distinct eigenvalues of $\hat{X}$ \emph{cannot} be assigned equal probability, and that if consistency is to be had, their probabilities must in fact be assigned according to the Born rule (but see section \ref{mn}). For example, for integer $m$, $n$ in the first instance, Deutsch can show that (his equation (3.5)/Wallace's Stage 3) \begin{equation} \mathcal{V} ( \sqrt{\frac{m}{n}} \vert x_1 \rangle + \sqrt{\frac{n - m}{n}} \vert x_2 \rangle  ) = \frac{m\mathcal{V}(c_1) + (n-m) \mathcal{V}(c_2)}{n}, \end{equation} and it is then fairly straightforward to generalise this result (Wallace's Stages 4 through 6) to the case of irrational coefficients and superpositions of arbitrary numbers of components, thus proving the Born rule. (See section \ref{egalitarianism} for more detailed discussion of this argument concerning unequal superpositions.)

 Given this result, we see that `rationality' in the sense of our axioms requires that the probability function appearing in the expected-utility rule must be the quantum amplitude-squared measure. \emph{This is a remarkably strong result.} In the standard case, Savage (1972) proved that \emph{the agent's preferences determine} a unique probability function (and an effectively unique value measure). However, there is substantial freedom over preferences within the bounds of rationality, and subsequently, in that case, substantial elbow room for rational agents to disagree over probability measures. (You and I can harbour very different subjective probabilities for rain tomorrow, even given the same evidence, without either of us being deemed irrational.) In the Everettian case, Deutsch and Wallace have proved that an agent is constrained to adopt precisely the amplitude-squared probabilities for action, given only that his preferences satisfy their constraining `rationality axioms': a set of decision-theoretic axioms, Physicality and Measurement Neutrality. (The agent's value function on Consequences, of course, suffers no such a priori restriction. The agent remains free to enjoy dancing in the rain.) The very strength of this result is likely to raise suspicion: just how can an agent whose preferences (say) disrespect amplitudes be branded `irrational'? I will argue later (section \ref{mn}) that this suspicion is well founded: it turns out that necessarily, such an agent violates Measurement Neutrality, but the intuitive justification for that axiom as a rationality axiom is not at all solid. First, though, I turn to a more sweeping worry, one that challenges the very coherence of the decision-theoretic approach: a worry as to whether decision theory is applicable to an Everettian measurement at all.

 \end{subsection}

 \end{section}

 \begin{section}{Problem: applicability of decision theory} \label{applicability}

 A prima facie problem with the decision-theoretic approach, as Wallace notes, is the following: since the relevant parts of decision theory are designed to deal with decision-making \emph{under uncertainty}, how can decision theory coherently be applied if determinism is the full Everettian story?

 To see the problem, consider again the standard decision-theoretic framework (section \ref{classicaldecisiontheory}): States, Consequences, Acts and Chance Setups. The point is that in a situation with no \emph{uncertainty}, it seems that much of this framework cannot apply. The set of States was supposed to be the locus of \emph{uncertainty}; the idea of a Chance Setup was that the agent was supposed to be \emph{uncertain} as to which State actually obtained; then the achievement of decision theory was to represent coherent preferences between Acts by means of a probability function that \emph{measures the agent's uncertainty} (and a value function on Consequences). If (classically) our agent knows the State of the world, it seems that all we can say to constrain his preferences is that they must define an ordering on Consequences. If that is true, then there is no point in defining an Act as a function from States to Consequences: we need only deal with one State (namely, the actual one). In particular, decision theory thus impoverished will have no nontrivial concept of probability: insofar as it deals in probability at all, it merely says that the actual State has probability 1, while all others have probability zero. It thus seems that \emph{even if} we accept that `probability', at least for current purposes, is defined by its role in rational action --- so, we accept that \emph{if} we advocated a strategy for rational action employing a certain quantity in the role of probability, that quantity would count as probability, no matter that it did not measure uncertainty --- \emph{still} we cannot have (nontrivial) probability without uncertainty in Everett.

 There are three lines of response to this problem. The first (Wallace's preferred response) is the \emph{subjective uncertainty (SU)} interpretation, which claims that although the Everett interpretation is \emph{objectively} a deterministic theory, an observer who believes Everett and is about to observe a QM measurement is nevertheless in a position of \emph{subjective uncertainty}. The second is the \emph{Reflection argument} - since a post-measurement observer who has not yet inspected her apparatus is in a position of genuine (self-locating) uncertainty and could be required to announce her betting quotient for the measurement outcome, a principle of reflection justifies applying decision theory to the pre-measurement observer. The third (my preferred response) is a \emph{fission-based interpretation}. This latter denies that uncertainty is a crucial component of the decision-theoretic approach at all, and applies decision theory on the basis that the pre-measurement observer cares about each of her future selves.

 In section \ref{su}, I argue that the subjective uncertainty argument fails --- that the result of an Everettian measurement, given the pure initial state, is (subjectively as well as objectively) \emph{certain}. The Reflection argument (section \ref{reflection}) suffers no such fatal flaw, but I will argue that, in comparison with the fission-based interpretation (section \ref{cf}), it obscures the real logic of the decision-theoretic programme.

 \begin{subsection}{The subjective uncertainty argument} \label{su}

 Consider Alice, observing a Stern-Gerlach experiment.\footnote{Or: consider Alice, about to undergo a classical brain-fissioning and transplant operation that will result in two successors, after which one of the successors will be given a card reading `spin-up' and the other a card reading `spin-down'. There is nothing particularly quantum-mechanical about this argument, other than its urgency.} Let us talk in terms of person-stages. Call the pre-measurement Alice `Alice$_1$'. Call the first post-fission copy of Alice `Alice$_2 ^{\rm up}$', and call the second post-fission copy of Alice `Alice$_2 ^{\rm down}$'.  How should Alice$_1$ view her$_1$ imminent splitting? Should she$_1$ feel as though she$_1$ knows what is going to happen, or should she$_1$ feel that she$_1$ is going to read either `up' or `down' on the dial but that she$_1$ does not know which?

 It has been argued (Saunders (1998)) that despite knowing all of the objective facts regarding her situation, Alice$_1$ should view regard her$_1$ \emph{subjective} situation as one of \emph{un}certainty --- despite knowing the objective truth that she$_1$ will split into two copies, one of which will observe each possible measurement result, she$_1$ should expect to become exactly one of those copies, and she$_1$ should feel uncertain as to which copy she$_1$ will become. In that case, the problem is solved --- decision theory can be applied from the point of view of the pre-measurement observer on the basis of her subjective uncertainty. I will argue, however, that the claim of subjective uncertainty should be rejected.

 \begin{subsubsection}{Common ground}
\label{commonground}

 My discussion of the (un)certainty issue will draw heavily on some fairly strong philosophical premises. These, however, form common ground between Saunders, Wallace and myself.

 First, we assume a reductionist account of personal identity over time. Second, we assume that the mental supervenes (somehow) on the physical. (The latter does not commit one to any very particular philosophy of mind, but only requires that if we have a complete physical description of a situation, there are no remaining undetermined parameters to decide questions of mentality.)

 These two assumptions will be crucial to the argument in the following way: without such philosophical commitments, the way is open to complain that in saying that before the fission there is one observer, and after the fission there are two observers each bearing the same causal and structural relations to Alice$_1$ that a person-stage normally bears to her very recent predecessors, we have not given the full story, since we have not explicitly stated which of the successors is the same person as Alice$_1$. Given our two assumptions, this need not be explicitly stated, since it is logically determined by the facts we \emph{do} explicitly state.

 Further, we agree on the description of this setup in terms of person-stages. There are three person-stages in the story: Alice$_1$, Alice$_2^{up}$ and Alice$_2^{down}$. Alice$_1$ sees a measurement apparatus in its `ready' state. Alice$_2^{up}$ sees an apparatus reading `up'; Alice$_2^{down}$ sees an apparatus reading `down'. Alice$_2^{up}$ and Alice$_2^{down}$ each bear the relation of personal-identity-over-time to Alice$_1$, in virtue of the causal and structural relations that obtain among the person-stages. (Alice$_2^{up}$ and Alice$_2^{down}$ do \emph{not} bear the relation of personal-identity-over-time \emph{to one another} --- the personal-identity-over-time relation between person-stages, if conceived as a symmetric relation, is not transitive.)

 The \emph{dis}agreement is over how we should connect this description in terms of person-stages to the issue of Alice$_1$'s (un)certainty.

 \end{subsubsection}

 \begin{subsubsection}{The argument for subjective uncertainty (SU)}
\label{susub}

 The SU claim is that a pre-fission agent who believes that her${\rm _1}$ future is \emph{objectively} certain should nevertheless treat an imminent fission event as \emph{subjectively uncertain}.

 To argue for this claim, Saunders (1998) considers the question of \emph{what Alice$_1$ should expect to see}, or, \emph{who Alice$_1$ should expect to become}. (As Saunders notes, these questions are equivalent, since it is exactly insofar as Alice$_1$ should expect to become Alice$_2 ^{\rm up}$ (Alice$_2 ^{\rm down}$ resp.) that she$_1$ should expect to observe spin-up (spin-down resp.) --- there is no doubt over what Alice$_2 ^{\rm up}$ and Alice$_2 ^{\rm down}$ observe.) Saunders considers a list of possibilities that he supposes to be jointly exhaustive, and argues against each alternative to advocate the SU view by process of elimination:

 (a) she$_1$ should expect nothing --- oblivion;

 (b) she$_1$ should expect to become both Alice$_2 ^{\rm up}$ and Alice$_2 ^{\rm down}$;

 (c) she$_1$ should expect to become exactly one of Alice$_2 ^{\rm up}$, Alice$_2 ^{\rm down}$, and should feel uncertain as to which she$_1$ will become.

 Saunders dismisses (a) as implausible --- double survival cannot constitute oblivion --- and I have no argument with this dismissal. But he then continues:

 \begin{quote} The genuine alternatives appear to be these: either she$_1$ anticipates being both [Alice$_2 ^{\rm up}$] and [Alice$_2 ^{\rm down}$], some kind of composite; or else she$_1$ anticipates being either [Alice$_2 ^{\rm up}$] or, in the exclusive sense, [Alice$_2 ^{\rm down}$].

 Nothing, both, or else just one of them? I have said that the first option is implausible \ldots As for the second option, it is straightforwardly inconsistent with the evidence; [Alice$_2 ^{\rm up}$] and [Alice$_2 ^{\rm down}$] do not speak in unison; they do not share a single mind; they witness different events. We do not know what it is to anticipate observing incompatible outcomes, at a single time. There remains only the third alternative: [she$_1$] anticipates being one of [Alice$_2 ^{\rm up}$] and [Alice$_2 ^{\rm down}$], but not both at once. (Saunders (1998), pp.383-4) \end{quote}

 So, Alice$_1$ expects to observe either spin-up exclusive-or spin-down, but \emph{doesn't know which}; hence subjective uncertainty.

 \end{subsubsection}

 \begin{subsubsection}{The argument for subjective certainty (SC)}
 \label{sc}

 The SC intuition is that if an observer-stage knows both the relevant aspects of the objective description of the universe, and his own location within that universe, there is no room for uncertainty, subjective or otherwise; and, further, that (due to the determinism of the Everettian quantum dynamics and our assumption that the pre-fission observer-stage knows the initial state) Everettian branching meets this condition.

 To give expression to this intuition, we need a way of connecting talk about the future to the basic person-stage account. I assume the following counterpart-theoretic account of that connection. For properties that we can assign to person-stages independently of considerations of personal identity over time, take as primitive the notion of a person-stage's having that property (say, the property of experiencing X). (We have already used this notion --- for instance, when we said that Alice$_2^{up}$ sees spin-up.) Then it is proper to say that Bob$_0$ \emph{will} experience X iff there is some future person-stage that both bears the relation of personal-identity-over-time to Bob$_0$ and experiences X. We can also talk, as Saunders does, of one person-stage's \emph{becoming} another: it is proper to say of a given person-stage Bob$_0$ that he \emph{will become} Bob$_1$ iff Bob$_1$ is a later person-stage and bears the relation of personal-identity-over-time to Bob$_0$. (This is proposed as an \emph{analysis} of locutions such as `will see X' and `will become Bob$_1$'.))

 Applying these notions to the case of Alice and her spin measurement, we get the following: the personal-identity-over-time relations among the person-stages are such that, according to our counterpart-theoretic account of talk of the future, Alice$_1$ will become Alice$_2^{up}$, and Alice$_1$ will become Alice$_2^{down}$. Similarly, Alice$_1$ \emph{will see} spin-up, and Alice$_1$ \emph{will see} spin-down.\footnote{This is not quite the full story. Later (section \ref{cf}), I will argue that in the quantum case, Alice$_2^{up}$ and Alice$_2^{down}$ each bear the relation of personal-identity-over-time to Alice$_1$ to degree less than unity. Thus I say that Alice$_1$ will see spin-up to some degree $p < 1$, rather than that she$_1$ will see spin-up \emph{simpliciter}. This point, however, does not affect the discussion of uncertainty.} 

 What of the crucial question: should Alice$_1$ feel uncertain? Why, Alice$_1$ is a good PI-reductionist Everettian, and she has followed what we've said so far. So she$_1$ \emph{knows} that she$_1$ will see spin-up, and that she$_1$ will see spin-down. There is nothing left for her to be uncertain about.

 What (to address Saunders' question) should Alice$_1$ \emph{expect} to see? Here I invoke the following premise: whatever she$_1$ knows she$_1$ will see, she$_1$ should expect (with certainty!) to see. So, she$_1$ should (with certainty) expect to see spin-up, and she$_1$ should (with certainty) expect to see spin-down. (Not that she$_1$ should expect to see \emph{both}: she$_1$ should expect to see \emph{each}.)

 Saunders will not be convinced by this argument: for him, it misses the point, by applying the wrong notion of expectation. The crucial point for him appears contained in the passage quoted above: `we do not know what it is to anticipate observing incompatible outcomes, at a single time.' That is, when we're talking of `expectation' in the sense of \emph{anticipation}, it is just a conceptual impossibility for Alice$_1$ to expect to see spin-up, and to expect to see spin-down, when no person-stage can see both. So he will reject my premise `whatever [Alice$_1$] knows she$_1$ will see, she$_1$ should expect (with certainty) to see.' The absurdity (by his lights) of my conclusion just shows (he will say) that we have a first-person notion of expectation that, in an Everettian situation, violates this premise. Now, looking for such a notion, we notice that there is a sense in which (I agree that) Alice$_1$ \emph{should} expect to observe spin-up or spin-down but not both: for \emph{each} of Alice$_1$'s individual successors-at-$t_2$, the statement `I observe spin-up or spin-down but not both' is true. Perhaps this is the key. One might further think that Alice$_1$ can feel \emph{certain} of just those experiences that \emph{all} her successors-at-$t_2$ have. Further, one might think that there is a fact of the matter as to whether or not Alice$_1$ will experience X at $t_2$ iff either all of Alice$_1$'s successors-at-$t_2$ experience X, or none of them do. If that is right, then in the Everettian scenario under consideration, there is no fact of the matter as to which outcome Alice$_1$ will see at $t_2$ (although it \emph{is} a fact that Alice$_1$ will see \emph{some} definite outcome at $t_2$); in this way uncertainty over what Alice will see can be reconciled with knowing all the facts that there are.\footnote{Thanks to Simon Saunders (in conversation) for explaining the points in this paragraph.}

 I disagree, in particular, with the claim that Alice$_1$ can feel certain of seeing X only if all her successors-at-$t_2$ see X. Although I agree that there is a sense in which Alice$_1$ expects to observe spin-up or spin-down but not both (namely, the sense outlined in the previous paragraph), I deny that \emph{this} sense delivers uncertainty. I say instead: \emph{I can feel uncertain over P only if I think that there is a fact of the matter regarding P of which I am ignorant.} (Note that this is a stronger condition than that it is \emph{not} the case both that there is a fact of the matter and that I know that fact; Saunders can accept this latter condition.) Otherwise I should feel certain --- certain that there is no fact of the matter, if I know that to be the case. (I assume eternalism as far as facts go --- that is, I take it that facts, in whatever sense they exist at all, exist timelessly, regardless of the spatiotemporal locations of the things in the world with which they are concerned. So I take it that, under a collapse interpretation of QM, there is (`already') a fact of the matter as to what the outcomes of future quantum experiments will be, no matter that that fact is unpredictable-in-principle from the present physical state of the universe; it is this that justifies my feeling uncertain under such an interpretation.)
 
 Now, if there is no fact of the matter as to whether P holds, that is (invariably, I think!) because enquiring as to whether or not P holds is a bad question --- a question with a false presupposition. For example, in quantum mechanics there will typically (in the absence of `measurement') be no fact of the matter as to exactly where an electron is located. But this means that `where is the electron?' is a bad question, not that we are to be uncertain as to where the electron is. (Do not assume that the electron has a location; ask instead, `what is its quantum state?') Similarly, there is no fact of the matter as to what Alice$_1$'s \emph{unique} successor-at-$t_2$ will see in a case of Everettian splitting. But \emph{if} it is such a fact we ask for when we ask `what will Alice see?', then this means that our question is a bad one, not that we are (or that Alice$_1$ herself is) to be uncertain as to what Alice$_1$ will see. (Do not assume that Alice$_1$ has a \emph{unique} successor;  ask instead, `what will \emph{each} successor-at-$t_2$ see?')

 We could, of course, take an operationalist line on (un)certainty, just as decision theory does on probability: we could say that to be uncertain as to whether or not one will observe X \emph{just is} to assign non-zero importance to each of the outcomes (observing X, observing not-X) when making decisions. In that case, it would be near-trivially true that the Everettian `feels uncertain'. But to make this move in defense of the applicability of decision theory would be viciously circular. If we were convinced in the first place that there was anything conceptually amiss with applying decision theory to a deterministic branching situation unless uncertainty can be fitted into that situation, we must have an independent notion of uncertainty --- independent, that is, of decision theory --- with respect to which we understand our problem. \emph{My} extra-decision-theoretic notion of uncertainty requires me to refrain from applying the term to anyone who thinks there is no relevant fact (objective or otherwise) of which she is ignorant. I conclude, therefore, that an Everettian facing an imminent quantum measurement has no right to feel uncertain.

 \end{subsubsection}

 \end{subsection}

 \begin{subsection}{The Reflection argument} \label{reflection}

 Despite my rejection of subjective uncertainty for the \emph{pre}-measurement observer, there \emph{is} a sense in which subjective uncertainty enters the Everettian picture --- each \emph{post}-measurement observer, before reading her apparatus, will have the subjective uncertainty of self-location. She will know that there is one branch containing an apparatus reading `up' and an observer-stage Alice$_2 ^{\rm up}$, and that there is one branch containing an apparatus reading `down' and an observer-stage Alice$_2 ^{\rm down}$, \emph{but she will not know which of these two observer-stages she, understood indexically, is}. Nevertheless, in our interpretation, the splitting \emph{has} occurred (by decoherence), and she \emph{is} determinately Alice$_2 ^{\rm up}$ or Alice$_2 ^{\rm down}$.

 How should this observer-stage regard her situation? As before, she knows all the objective facts. But now, there is a subjective fact of which she is ignorant, namely, her own identity. (That is, she is ignorant of a `de se' fact in the sense of Lewis (1979).) It seems clear that Alice$_2 ^{\rm up}$ and Alice$_2 ^{\rm down}$ should each regard \emph{their} situations as ones of \emph{subjective uncertainty}.

 So far, though, we have been discussing a \emph{post}-fission observer. We have said nothing about the \emph{pre}-fission observer, whose rational behaviour we seek to derive. We cannot invoke an \emph{epistemic} reflection principle (such as that introduced by van Fraassen (1984)) to argue that the pre-measurement observer should also be uncertain --- the post-measurement uncertainty is self-locating, and the indexical facts over which Alice${\rm _2 ^{up}}$ and Alice$_2 ^{\rm down}$ are uncertain are simply inapplicable to Alice${\rm _1}$. But now consider the following \emph{decision-theoretic reflection principle}:

 \begin{quote} If, at time $t$, I decide rationally to pursue a certain strategy at a later time $t'$, and if I gain no new information relevant to that strategy between times $t$ and $t'$, then it is rational [i.e. rationally compelling] not to change my choice of strategy at $t'$. (Wallace (2002b), p.58) \end{quote}

 This principle, I think, is a reasonable one, and it holds the key to the second line of response to the applicability problem: if we want to apply a decision theory whose application is justified under uncertainty, we can apply it directly to the post-fission observer-stages on the basis of their genuine uncertainty, and appeal to reflection in order to argue that the pre-fission observer-stage should adopt the same strategy for action.

 The Reflection argument achieves its purpose. My main objection to it is that by my lights, it obscures the real logic of the argument, which latter is manifest in the fission interpretation. I will elaborate on this point after presenting the latter approach (section \ref{cf}).

 \end{subsection}

 \begin{subsection}{The fission interpretation} \label{cf}

 Parfit (1984) argued that what matters in survival --- and therefore, presumably, what matters in \emph{quality} of personal future also --- is the existence of future person-stages bearing appropriate relations of structural similarity and causal connectedness to one's present person-stage. This dictates that in a case of fission, I-now must care about each appropriately related successor as one of my future selves - the rationale being that, whether we've realized it or not, this is in any case all there has ever been to caring about our futures. This being so, we should be able to develop prescriptions for rational action that altogether bypass the issue of uncertainty. The task of the present subsection is to argue that decision theory (with reasonable interpretational adjustments) can do just that, and that \emph{this} is the way to understand probability in Everett.

 Recall the problem: it seems that without uncertainty, several of the decision-theoretic axioms are simply inapplicable, and so there must be something seriously conceptually amiss with a proof (such as Deutsch's) that claims to establish probability by making essential use of those axioms.

 The fission-based solution is as follows: uncertainty is not after all an indispensable prerequisite to the application of decision theory. Rather, the knowledge that one will undergo fission can do all the work conventionally done by uncertainty over the world's State. To put it in a way that brings out the analogy: the knowledge that one will have many actual successors can do the work conventionally done by the prospect of many possible successors. In either case our agent must consider the interests of all successors: in the case of uncertainty, because he cares about his unique future self but does not know which successor is to be real, and in the case of fission, because all successors are real, and he cares about them all as his future selves. So we can define an Act as a function from States of a \emph{branch} to Consequences, no matter that there is no uncertainty over whether a branch in that State will exist, or over whether the successor in that branch \emph{is me}. Our problem (in the first two paragraphs of section \ref{applicability}) arose under the assumption that we must treat a future macroscopic superposition of successors as an \emph{atomic} entity for decision purposes, that we could not break it down and require preferences to recognize certain salient features of the superpositions (such as that a certain component of the superposition represents a successor of amplitude $a_i$ enjoying a Consequence $C_i$). If that were the case, then indeed, we could constrain the agent's preferences only by requiring an ordering on the space of all possible future superpositions, and no nontrivial probability could arise. But the assumption is unjustified.

 Let us see what work decision theory will be doing in such a fission context. The Parfitian account of caring for the future tells me to care about all my successors. What it \emph{doesn't} tell me, at least not directly, is how to weigh the interests of my successors \emph{one against another}, or how to choose between Acts that would generate different numbers of successors. This will be crucial in determining my rational actions. Suppose (somewhat artificially) that I am confronted with the choice of whether or not to sign a particular form. After I make my choice, a Stern-Gerlach experiment will be carried out. If I signed, my spin-up successor will be force-fed olives, and my spin-down successor will be given chocolate. If I didn't sign, no-one is fed. But I hate olives. How do I decide whether having one successor worse off and another better off is preferable to having all my successors stuck with the status quo? Do I sign or not?

 Lewis (1976) introduces the term `R-relation' to represent the thing he considers matters in survival: mental continuity and connectedness. As Lewis recognizes, the holding of the R-relation between given person-stages is a matter of degree: for any two given person-stages, he suggests, a real number $r \in [0,1]$ represents their degree of R-relatedness. Ordinary personal identity over an infinitesimal time interval involves person-stages R-related to degree 1, while person-stages of entirely disjoint continuant persons are R-related to degree 0; cases of fading memory, fission and fusion, brainwashing, brain transplant and so on may take intermediate values.

 I agree that the R-relation --- mental continuity and connectedness --- captures much of what matters in survival. But I do not wish to beg the question by assuming that this is \emph{all} that matters; indeed, I will argue that given the Everett interpretation and Deutsch's axioms, it is \emph{not} all that matters. I therefore introduce the term `R*-relation', initially as a place-holder, to denote \emph{whatever it is that matters in survival and in caring for the future}.

 The question is to what extent the R*-relation holds between myself and each of my Everettian successors. It is here that decision theory claims to help. Where, classically, decision theory invokes a notion of \emph{Chance Setups} (recall: `situations in which a number of possible events might occur, and \ldots it is in general impossible for the agent to predict with certainty which will occur'), we can invoke \emph{Fission Setups} (situations in which the agent will undergo fission, and a number of distinct successors will be correlated with qualitatively distinct events). Where classical decision theory construes Acts as `possible courses of action for the individual (usually in the face of uncertainty as to which outcome is going to occur)', and the set $\mathcal{S}$ of States as the locus of uncertainty, we can construe Acts as possible courses of action for the individual, usually in the face of knowledge that the individual will have multiple successors, each in general correlated with qualitatively different States (and hence Consequences). No formal adaptation, and no further interpretational adaptation is required --- Deutsch's and Wallace's decision-theoretic axioms can be applied as well, and as reasonably, under a fission-based interpretation of the decision structure as under its uncertainty-based analogue. The Deutsch-Wallace proof is to be understood as claiming to establish that my Everettian successors bear the R*-relation to me-now in proportion to their amplitude-squared measures. With this conceptual base, decision theory can be applied without uncertainty.

 We would then have precisely what I have argued (section \ref{theproblem}) is needed to solve the probability problem in Everett: a rationality principle that tells us how to act in the face of the Everettian world-picture, and that (arguably) instructs us to regard Everett as empirically confirmed.

 I can now say why, in my view, the Reflection approach obscures the logic of Everettian decisions. Suppose, for the sake of argument, that, in the above scenario, the relevant outcomes (chocolate, hunger, olives) are equally spaced in terms of utility. Then if I strictly prefer signing the form to not signing it (perhaps in a fission setup in which the spin-down successor has higher quantum-mechanical weight), that is so \emph{because} I care more about my spin-down successor than I do about my spin-up successor. It is not \emph{because} my post-fission selves will suffer weighted disorientation, although (as emphasized by the Reflection argument) that is also true.

 \end{subsection}

 \end{section}

 \begin{section}{Problem: justification of measurement neutrality (MN)} \label{mn}

 \begin{subsection}{Measurement neutrality} \label{mnsub}

 My discussion up to now has been conditional --- \emph{if} we accept the decision-theoretic axioms, Physicality (P) and Measurement Neutrality (MN), and the applicability of decision theory to deterministic fission, \emph{then} the desired result follows.\footnote{
 It \emph{does} follow: the proof, once Deutsch's tacit assumptions have been made explicit as Wallace urges, is valid. Further, Deutsch's tacit assumptions \emph{are not themselves directly probabilistic in nature}, so we need not take them to undermine the motivation for the decision-theoretic project. See Wallace (2003a) for a reply to the contrary claim made by Barnum, Caves, Finkelstein, Fuchs and Schack (2000): the point is that we are to understand Deutsch's proof as applying to the Everett interpretation, in which measurement is treated as a quantum-mechanical process, rather than taken as primitive.} I have argued (section \ref{cf}) in defense of the decision-theoretic axioms, and I regard Physicality as obviously correct. This leaves one question, whose answer will determine whether we are to regard the Deutsch-Wallace theorem as a genuine \emph{proof} of the Born rule, or as an interesting conditional with a dubious antecedent: is the assumption of Measurement Neutrality justified?

 Measurement Neutrality, recall (from section \ref{quantumrepresentationtheorem}), is the assumption that a rational agent should be indifferent between any two quantum games that agree on the state $\vert \psi \rangle$ to be measured, measurement operator $\hat{X}$ and payoff function $\mathcal{P}$, regardless of \emph{how} $\hat{X}$ is measured on $\vert \psi \rangle$. It is this assumption that justifies the abstraction from physical detail involved in Deutsch's and Wallace's notation (i.e. the representation of quantum games (Acts) by \emph{triples} $\langle \vert \psi \rangle, \hat{X}, \mathcal{P} \rangle$, rather than by \emph{quadruples} $\langle \vert \psi \rangle, \hat{X}, \mathcal{P}, \omega \rangle$, where $\omega$ would represent additional detail of the physics): according to MN, a rational agent will always be indifferent between two quadruples $\langle \vert \psi \rangle, \hat{X}, \mathcal{P}, \omega \rangle$ and $\langle \vert \psi \rangle, \hat{X}, \mathcal{P}, \omega ' \rangle$, and so we need not distinguish the two.

 At first sight, this assumption seems so obviously correct as hardly to be worth a second glance (`who cares exactly \emph{how} a measurement device works, provided that it works?'). As Wallace (2003b, pp.21-2) points out, however, this is not the case. To see why, we can consider the conflict between Measurement Neutrality and a Born-violating rationality strategy that I will call `Egalitarianism'; this is the task of the next section.

 \end{subsection}

 \begin{subsection}{Measurement Neutrality versus Egalitarianism} \label{egalitarianism}

 Historically, when it has been supposed that probability can make sense at all in the Everett interpretation, it has often been asserted (e.g. Graham 1973, p.236) that, for an Everettian, each branch must receive \emph{equal} probability --- after all, each branch is equally real. When (as now) probability is to be understood in terms of rational caring, a similar point can be pressed: since all my successors are equally real, surely I should care for each to an equal degree? Let `Egalitarianism' denote the view that indeed, I should care equally for each.

 To see just how non-trivial MN is, consider how it parts company from such Egalitarianism in the following case (which is, in fact, integral to Deutsch's proof). One way to measure an observable $\hat{X}$ on a system A in a state of superposition $\vert \psi \rangle {\rm _A} = a_1 \vert x_1 \rangle {\rm _A} + a_2 \vert x_2 \rangle {\rm _A}$ is first to couple that system to an auxiliary system B, on which an observable $\hat{Y}$ with eigenspectrum $\vert y_i \rangle$, $i \in \{1, ..., N\}$ is defined, in such a way as to generate the following entangled state:

 \begin{equation} \vert \psi \rangle {\rm _{AB}} = \frac{a_1}{\sqrt{n}}\vert x_1 \rangle {\rm _A} ( \vert y_1 \rangle {\rm _B} + \ldots + \vert y_n \rangle {\rm _B}) + \frac{a_2}{\sqrt{N-n}} \vert x_2 \rangle {\rm _A} (\vert y_{n+1} \rangle {\rm _B} + \ldots + \vert y_N \rangle {\rm _B}) \end{equation}

 We can then measure $\hat{X}$ on A by measuring $\hat{Y}$ on B, and counting the observation of any $y \in \{y_1,  \ldots, y_n\}$ on B as an observation of eigenvalue $x_1$ on A, and any $y \in \{y_{n+1},  \ldots, y_N\}$ on B as an observation of eigenvalue $x_2$ on A. But a second way to measure $\hat{X}$ on A, of course, is to dispense with the auxiliary system altogether and make the measurement of interest directly.

 How should the rational agent, who (say) has already accepted a deal under which the eigenvalue $x_1$ of $\hat{X}$ is to be correlated to some Consequence $c_1$ and $x_2$ to a Consequence $c_2$ --- so, he has already accepted the $\vert \psi \rangle$, $\hat{X}$ and $\mathcal{P}$ that for Deutsch and Wallace constitute the full specification (for decision purposes) of a quantum game  --- feel about a choice between these two ways of measuring $\hat{X}$? Measurement Neutrality requires him to be indifferent. But, on the picture entertained by the Egalitarian, all is far from the same --- the branching structures associated with the two measurements are very different. On that picture, for example, a direct measurement of $\hat{X}$ on A will result in two branches, one realizing $x_1$ and the other realizing $x_2$, whereas the above measurement of $\hat{X}$ on A via coupling to B will result in $N$ branches, in $n$ of which $x_1$ is realized, in $(N- n)$ of which $x_2$ is realized. In general, then, an agent who cares about how many successors he has, or about what proportion (by simple counts) of his successors receive each Consequence, will \emph{not} be indifferent, therefore such an agent will violate MN. And these do not (at first sight) seem unreasonable things for an Everettian to care about; MN does not \emph{only} require indifference between measuring devices with some concealed component painted blue and otherwise identical devices in which the component is painted red.\footnote{It may be possible to give MN a far stronger intuitive justification from an SU, rather than a fission-based, perspective. (See Wallace (2003c; 2003b, p.22) for some argument in this direction.) If so, that may give a reason (for an Everettian!) to wish SU true. But a reason to wish SU true is not a reason to believe it true; I have argued (section \ref{su}) that it is false.}

 At what point in Deutsch's proof should an agent who cares about the branching structure in this way feel that injustice is being done? Technically, of course, the proof cannot even get off the ground without Measurement Neutrality: as mentioned above, MN is built in right from the start, being incorporated into Deutsch's notation. But heuristically, it would be nice to identify the precise point at which the disagreement really sets in: Deutsch's equations (2.10) and (3.1) (Stages 1 and 2 in Wallace's reconstruction of Deutsch's proof) --- equiprobability in case of equal amplitudes --- are unobjectionable to our Egalitarian. Not so Deutsch's equation (3.5) (Stage 3 in Wallace's reconstruction), which claims that the agent is constrained to assign amplitude-squared probabilities to (certain cases of) \emph{unequal} amplitudes: our agent wants to assign \emph{equal} probabilities to each of his successors, regardless of amplitude. Unsurprisingly, it turns out that Measurement Neutrality plays a crucial role in deriving Deutsch's (3.5). Deutsch's strategy is to argue that:

 \begin{itemize}

 \item It has already been proved that, if we accept the axioms listed above, we must assign equal probabilities in case of equal amplitudes.

 \item Therefore, in the scenario described above, we must assign equal probabilities to each branch that corresponds to a distinct result for the measurement of $\hat{Y}$ on system B.

 \item But the same physical procedure can be regarded as a measurement of $\hat{X}$ on system A: we can count the observation of any $y \in \{y_1,  \ldots, y_n\}$ on B as an observation of eigenvalue $x_1$ on A, and any $y \in \{y_{n+1},  \ldots, y_N\}$ on B as an observation of eigenvalue $x_2$ on A.

 \item Therefore, when the measurement of $\hat{X}$ is conducted in this way, we must assign unequal probabilities (with particular values) to the possible results for $\hat{X}$.

 \item By Measurement Neutrality, we must assign those same unequal probabilities to the possible results for $\hat{X}$ when $\hat{X}$ is measured in any \emph{other} way (including methods in which the auxiliary system is dispensed with and the measurement is made directly).

 \end{itemize}

 Our Egalitarian should object at the last step of this argument: he should say that, for sure, \emph{if} $\hat{X}$ is to be measured in such a way that the numbers of branches corresponding to the results $x_1$, $x_2$ are in the ratio $\frac{\vert a_1 \vert ^2}{\vert a_2 \vert ^2}$, \emph{then} he will assign probabilities $\vert a_1 \vert ^2$, $\vert a_2 \vert ^2$ to $x_1$, $x_2$ respectively, but that if $\hat{X}$ is measured in a way that generates one branch for each of $x_1$, $x_2$, then he will assign probability $\frac{1}{2}$ to each. Neither will he be impressed by any foot-stamping insistence that he is \emph{unreasonable} to care about how his observable is measured: as I have argued, such insistence merely begs the question against his view.\footnote{A similar point applies to Wallace's proof from Savage-style decision axioms: the Egalitarian will object to the use of Measurement Neutrality in proving Wallace's General Equivalence Theorem (Wallace (2002b, p.34-5/2003b, p.7-10)).}

\end{subsection}

\begin{subsection}{Egalitarianism is incoherent}
\label{failureofegalitarianism}

 Egalitarianism is not, in fact, a tenable position. The reason is that it relies on an over-idealized conception of Everettian branching: it relies on the assumption that the `number of branches' associated with a given outcome is well-defined. This assumption is common enough --- it underwrites the most natural turns of phrase for discussing Everettian branching, which latter I have used throughout this paper --- and, \emph{if used with caution}, harmless enough. But \emph{it is not generally true}, and here the failure to keep that in mind is taking its toll.

 Why is the number of branches associated with a given outcome not well-defined? The point is that we are assuming a \emph{decoherence-based} conception of `branching', not a primitivist conception. Under any such conception, the `number of branches' at a given time could only be the number of elements of the decoherence-preferred basis that appear in the universal state with nonzero coefficient at that time. But, in the first instance, the decoherence basis is only approximately defined by the dynamics; decomposing the universal state into branches by projections onto a slightly rotated basis can wildly affect the \emph{number} (although not the total weight) of branches enjoying a given Consequence. Second, there is ambiguity over just how \emph{coarse-grained} the division of the multiverse into distinct branches is to be: too fine-grained a division will fail to retain the freedom from inter-branch interference that played a crucial role in selecting the preferred basis in the first place; too coarse-grained a division will lead to loss of macroscopic definiteness; we must choose some compromise, but who is to say \emph{exactly} what degree of coarse-graining we should settle on? Again, our exact choice can dramatically affect relative numbers of successors enjoying each Consequence, but it is an arbitrary choice, and there is no deep fact of the matter. Without abandoning the whole philosophy behind the decoherence approach --- which would be a desperate expedient --- there is no way to fix one \emph{exact} basis, or degree of coarse-graining, as `preferred' for purposes of rational decision. Third, even if (\emph{per impossibile}) these problem could be solved, branching will in general be \emph{continuous} --- for instance, many of the interactions that lead to branching involve monitoring a system's \emph{position}, and position is a continuous variable. We cannot, therefore, \emph{properly} talk of `how many successors' see spin-up, and Egalitarianism fails.\footnote{Wallace (2003c), section 8, provides a more extended presentation and defense of this important point.}

\end{subsection}

 \begin{subsection}{The status of the Everettian Born rule without uncertainty}
\label{spectrum}

 There is a spectrum of possible positions for an Everettian who accepts the Born rule. At one extreme, she could claim that a compelling argument \emph{forces} any rational agent to accept the Born rule, on pain of denying some obviously true premise. At the other extreme (primitivism), she could simply \emph{stipulate} that rational agents are to accept the Born rule. Or she could take an intermediate position: she could go beyond \emph{mere} stipulation to offer \emph{some} supporting arguments that, however, fall short of absolute compulsion.

 In this section, I want to make three claims. (i) (Section \ref{primitivism}) Primitivism is, indeed, plausible. This provides the Everettian with, at the very least, a `fall-back' position --- while she may hope for more substantial justification for the Born rule, she need not fear that without such additional justification, her interpretation is sunk. (ii) (Section \ref{alternatives}) We can do \emph{somewhat} better than primitivism, by noting that it turns out to be extremely difficult to find any plausible rationality strategy that violates the Born rule. (iii) (Section \ref{takingstock}) Because of a lack of justification for Measurement Neutrality, the Deutsch-Wallace decision-theoretic proof does not (under the present interpretation) play the persuasive role its originators hoped for: we cannot do \emph{all that much} better than primitivism.

 \begin{subsubsection}{Primitivism is not so bad}
\label{primitivism}

 \emph{Ceteris paribus}, the existence of a compelling argument to force the Born rule would improve the Everettian position. A philosophical position is always improved if a principle it formerly accepted as primitive can be shown to follow from other accepted primitives, or from the otherwise obviously true. But this does not have to mean that the \emph{absence} of a compelling argument should lead us to abandon the idea that the Born rule governs Everettian rationality --- primitivism may suffice. (This point has been pressed, in particular, by Papineau (1996), who points out that insisting on a compelling argument for Everettian probability would be invoking double standards --- what argument forces us, for instance, to consider \emph{propensities} as relevant to rationality?)

 The primitivist suggestion amounts to this: if we have to accept the \emph{conclusion} of the quantum representation theorem (as the Fission interpreter understands it) as primitive, viz. that a rational Everettian is to care about her successors in proportion to their amplitude-squared measures, this is not, in any case, so bad.\footnote{Similar suggestions --- specifically, that the quantitative problem of probability in Everett can be solved by \emph{fiat}, by simply \emph{stipulating} that the amplitude-squared measure is to be interpreted as `probability' --- have been made before. But without the link to rationality, it is just not clear what such suggestions could possibly \emph{mean} in an Everettian picture. Accepting some suggestion as primitive is one thing; accepting as primitive a suggestion \emph{that has no clear meaning} is quite another.}

 I make this point by means of an analogy with another dilemma that, I take it, we have (or, at least, the Everettian has) already learnt to live with: the potential for a reductionist account of personal identity over time to undermine self-interest. This analogy is set out in the following table.

{\vspace{6mm}}

 \begin{tabular}{p{2.2in}p{2.2in}}

 \textbf{Self-interest and reductionism} & \textbf{Rationality and Everett} \\

{\vspace{0mm}}Under a primitivist conception of personal identity over time (PIOT), I (selfishly) care about my future self \emph{because it's me}. This seems intuitively acceptable enough. &

{\vspace{0mm}}Under (say) a primitivist conception of probability, I am more concerned to maximize the utility that's correlated with high-probability outcomes \emph{because they're more likely to be real}. This (perhaps!) seems intuitively acceptable enough. \\

{\vspace{0mm}} Under a reductionist account of PIOT, this becomes: I am to (selfishly) care about future person-stages that bear certain structural and causal relations to me-now. &

{\vspace{0mm}}Under an Everettian picture, this becomes: I am more concerned to maximize my utility in high-weight worlds.

 \end{tabular}

 \begin{tabular}{p{2.2in}p{2.2in}}

 \textbf{Self-interest and reductionism} & \textbf{Rationality and Everett} \\

{\vspace{0mm}}This invites the challenge: what makes these particular relations relevant to self-interest? &

{\vspace{0mm}}This invites the challenge: what makes amplitudes relevant to rationality?\\

{\vspace{0mm}}Answer: that's just what self-interest is (whether or not we've realized it). &

{\vspace{0mm}}Answer: that's just what rationality is (whether or not we've realized it).\\

{\vspace{0mm}}The challenge may be pressed: why is `self-interest' in this sense prescriptively compelling? &

{\vspace{0mm}}The challenge may be pressed: why is `rationality' in this sense prescriptively compelling?\\

{\vspace{0mm}}This question seems to have no answer other than an evolutionary one: person-stages that failed to look out for `their' futures in this way died out, and so don't have any present structural counterparts. &

{\vspace{0mm}}This question seems to have no answer other than an evolutionary one: person-stages that failed to prioritize their future \emph{high-weight} structural counterparts don't have present structural counterparts \emph{in high-weight branches}. \\

{\vspace{0mm}}It's perfectly consistent for a person-stage to refuse to care about future structural counterparts: sure, he will tend to die out, but he doesn't care. &

{\vspace{0mm}}It's perfectly consistent for a person-stage to refuse to care about weight: sure, he will tend to die out in high-weight branches, but he doesn't care. \\

{\vspace{0mm}}However, the self-consistency of such lack of care doesn't undermine the prescriptive force of acting `self-interestedly' in this sense.&

{\vspace{0mm}}However, the self-consistency of such lack of preferential care doesn't undermine the prescriptive force of acting `rationally' in this sense. \end{tabular}

{\vspace{6mm}}

 The suggestion, then, is that the rational Everettian \emph{just should} care about her successors in proportion to their amplitude-squared measures. How can we talk, to bring out the reasonableness of this suggestion? We had better not say that the higher-weight branches contain \emph{more copies} of us --- as noted above (section \ref{failureofegalitarianism}), the number of branches is not defined. But since we have a measure over our successors, we can, if we find it intuitive, talk of `how much successor' sees spin-up. I have a preference for my spin-down successor to receive chocolate, rather than my spin-up successor, because there is \emph{more of} the former; more of my future lies that way. Thus, I think, Lockwood's (1996) talk of a `superpositional dimension', and/or Vaidman's (1998, 2001) suggestion that we speak of the amplitude-squared measure as a `measure of existence', are somewhat appropriate (although we are not to regard lower-weight successors as \emph{less real}, for \emph{being real} is an all-or-nothing affair --- we should say instead that there is \emph{less of} them).

 However that may be, any reader who dislikes such talk is welcome to stick with the following. There is some relation --- I have (section \ref{cf}) called it the R*-relation --- that my successors/classes of successor(s)/etc. bear to me to varying degrees according to their amplitude-squared measures. That relation can be taken as holding between branch-stages (i.e. world-stages) rather than person-stages, if we want to avoid an apparent emphasis on self-interest. We may think, as I urge in the above table, that that relation governs rational care-for-the-future in the manner given by the Born rule; if so, no more needs to be said.

 \end{subsubsection}

 \begin{subsubsection}{Improving on primitivism: the absence of sensible alternatives to the Born rule}
\label{alternatives}

 The failure of Egalitarianism (section \ref{egalitarianism}) raises an important question: are there \emph{any} coherent rationality strategies that are at all plausible and that also violate the Born rule? If not, we may hope to go beyond primitivism, and to defend the Born rule by sheer process of elimination.

 It's certainly true that MN-violating strategies that guide an agent's actions \emph{other than in an ad hoc manner}, while remaining faithful to Physicality, are extremely difficult to come by --- far more difficult than one would initially suspect. Suggestions that probabilities could be set equal to \emph{squared} branch weights (renormalized), for instance, likewise fall foul of the requirement that advice for rational action should be unaffected by slight rotations of the preferred basis or alterations of the degree of coarse-graining, and of the failure of the branching structure to be discrete. Suggestions that \emph{eigenvalues}, for instance, might be relevant to probabilities, fall foul of Physicality even \emph{without} invoking the approximate nature of the preferred basis: the eigenvalue attached to a given measurement outcome is an artefact of our description, and is not determined by the physics. Barnum \emph{et al} (2000) complain that Deutsch's stated axioms fail to outlaw probability prescriptions such as `the result associated with $\vert \phi \rangle$ \emph{always} occurs'; whereas Barnum \emph{et al} take this to illustrate that Deutsch needs a `hidden \emph{probabilistic} assumption' (emphasis added), the suggested rule actually violates Physicality (Wallace (2003b), section 6). It does seem that the only way we can come up with an MN-violating strategy is by brute force: that is, by specifying a preference ordering over Acts on a case-by-case basis, without appeal to any general governing rationale.

 What's wrong with that? Perhaps, we might think, nothing --- after all, an arbitrary preference ordering over possible states of the multiverse would provide a guide to rational action that is at least \emph{self-consistent} and \emph{well-defined}; there is no violation of Physicality here. It will not in general be a preference ordering that we can regard as intuitively compelling, but no matter: after all, \emph{classical} decision theory allows the `rational' agent total freedom to hold a preference ordering that most of us would consider bizarre.\footnote{This point is due to Adam Elga (in conversation).} If we are to rule out such arbitrary preference orderings, it must be on grounds stronger than mere self-consistency and Physicality.

 However, I think that imposing such stronger grounds is legitimate. According to my argument of section \ref{cf}, the correct classical analog of this arbitrary preference ordering over wavefunctions of the multiverse is an arbitrary preference ordering over \emph{Chance Setups}, not over Consequences. Any such ordering would be perfectly self-consistent --- but, in general, it would violate the decision-theoretic axioms. Even in the classical case, therefore, we are in the habit of imposing rationality requirements stronger than consistency, and we regard that practice as acceptable just in case the axioms themselves seem intuitive enough, and lead to intuitively acceptable enough prescriptions for action. The imposition of the Born probability measure is just such a move.

 \end{subsubsection}

 \begin{subsubsection}{Taking stock: The role of decision theory in a defense of the Everett interpretation}
\label{takingstock}

 Where does all this leave the decision-theoretic programme?

 Deutsch and Wallace claimed to be offering a compelling argument --- viz., the decision-theoretic proof --- that would force any rational agent to accept the Born rule, on pain of rejecting some supposedly undeniable premises. Given the interpretation urged in this paper, I think this is not quite the right perspective on the proof. According to the Everett interpretation, `measurement' is just another physical process; it is therefore extremely hard to see \emph{directly} why equivalence \emph{qua} measurement should force equivalence \emph{qua} chance setup. (It is important here to note that Egalitarianism failed \emph{due to its own incoherence}, not due to any compelling nature of MN.) In particular, \emph{if} some rationality strategy could be found that seemed \emph{otherwise} intuitively compelling, to invoke Measurement Neutrality against that strategy would be merely to beg the question.

 Rather, I think the correct perspective is the following. The Everettian Born rule is fairly compelling once we see (a) that the competing genuine intuition (Egalitarianism) is incoherent, and (b) that it is actually extremely difficult to cook up an alternative rationality strategy that respects Physicality and is also systematic. Perhaps combined with a hint of primitivism, these considerations should lead us to accept that, in an Everettian world, the preferences of rational agents over quantum Acts would be such that they can be represented by a utility function over Consequences and the Born probability measure over measurement outcomes. The Born rule, of course, implies Measurement Neutrality, so \emph{this} should lead us to accept the latter. (So my point is not that MN is \emph{false} --- I think it is true. My point is that MN is unacceptable as a \emph{primitive}.) 

 We might think that, if this is right, the decision-theoretic program is simply left out in the cold: after all, if we are content to accept the Born rationality strategy as more-or-less primitive, what use do we have for a proof that it follows from a principle (MN) that we \emph{don't} accept as primitive? This, though, would be a little too negative. 
 
 First, decision theory (as in the classical case, and quite apart from the Deutsch-Wallace theorem) provides an appropriate and highly illuminating structure for discussions of probability, demonstrates the existence of a probability measure or `caring measure' over future branches, and makes it clear what is meant by that measure, by connecting that measure to a preference ordering over Acts.
 
 Second, having accepted the Born rule and seen that it implies Measurement Neutrality, we can (if we wish) run the Deutsch-Wallace proof to see that Measurement Neutrality (together with thoroughly reasonable decision-theoretic axioms and the uncontentious Physicality) in turn implies the Born rule; the Deutsch-Wallace proof then serves as an illustration of the internal coherence of the recommended Everettian position on probability. 
 
 Third, we may hope to do better. It may be possible to find an intuitive justification for an intermediate primitive --- stronger than MN, but weaker than the full Born rule --- and thence to recover the full importance of the decision-theoretic proof as a \emph{persuasive} argument for the Everettian Born rule. Wallace (2003c, section 8) suggests one such intermediate primitive (`Equivalence'): it suffices if we assume (in addition to Physicality and decision theory) that \emph{the agent assigns the same probability to any two Consequences that have the same quantum-mechanical weight}. I have not embraced this path, because I find it hard to see how this assumption can be intuitively justified other than by relying on the intuitive plausibility of the Born rule itself. But perhaps there is a way.

 \end{subsubsection}

 \end{subsection}

 \end{section}

 \begin{section}{Conclusions} \label{conclusions}

The problem of probability in Everett is often formulated as a twofold challenge: an incoherence problem (how can probability make sense, if all outcomes are realized?) and a quantitative problem (assuming probability makes sense at all, why the particular probabilities prescribed by the Born rule?). Under the present approach, the incoherence problem in particular is transformed. We note first that merely verbal disagreements need not detain us: I have argued (section \ref{theproblem}) that the Everettian has no need to claim title to the term `probability', over and above her needs (a) to formulate a strategy for rational action in the face of branching, and (b) to be entitled to regard quantum mechanics, given the sequences of experimental outcomes we have in fact observed, as empirically confirmed. If it can be shown that a certain rationality principle holds (viz. that the preferences of rational agents over quantum games are represented by an expected utility rule that uses the Born probability measure), this will meet (a) and, I have suggested (section \ref{link}), there is reason to hope that it may also meet (b). If that is right, then the problem of probability in Everett reduces to the problem of how to justify the requisite rationality principle.\footnote{The following objection is sometimes raised against the decision-theoretic approach: in an Everettian context, all outcomes of a decision are realized, and therefore it simply does not make sense to make choices, or to reason about how one should act. If that is correct, then while we may agree that probability can in principle be derived from rationality, this is of no use to the Everettian, since (it is claimed) the Everettian cannot make sense of rationality itself.

If this was correct, it would be a pressing `incoherence problem' for the decision-theoretic approach. The objection, however, is simply mistaken. The mistake arises from an assumption that \emph{decisions} must be modelled as Everettian branching, with each possible outcome \emph{of the decision} realized on some branch. This is not true, and it is not at all what is going on in the decision scenarios Deutsch and Wallace consider. Rather, the agent is making a genuine choice between \emph{quantum games}, only one of which will be realized (namely, the chosen game). To be sure, each game consists of an array of branches, all of which will, \emph{if} that game is chosen, be realized. But this does not mean that \emph{all games} will be realized. It is no less coherent for an Everettian to have a preference ordering over quantum games than it is for an agent in a state of classical uncertainty to have a preference ordering over classical lotteries.}

The decision-theoretic approach attempts to meet precisely that challenge: to provide a rigorous argument to force the Everettian Born rule as a rationality principle. This approach, however, requires us (in the first instance) to apply decision theory to Everettian branching, \emph{with decision-theoretic Consequences assigned to individual branches} (rather than simply to states of the multiverse). We may well ask why this is reasonable: here, Wallace (2002b, 2003b) has advocated an appeal to subjective uncertainty. I have rejected this appeal to uncertainty (section \ref{su}), but I have proposed and defended an alternative (section \ref{cf}): the Everettian is to care about each of her multiple futures in the same way (and for the same reasons) that anyone might care for a unique future, and \emph{this} licenses application of analogs of the usual decision-theoretic axioms to quantum branching in the required way --- uncertainty is not required.

Deutsch and Wallace also need to place two further restrictions on the `rational' agent's preferences over quantum games: Physicality and Measurement Neutrality. Their proofs then establish that \emph{if} an agent's preferences over quantum games satisfy all of these constraints, \emph{then} not only do his preferences (as in classical applications of decision theory) define a unique probability function, but (a restriction that has no classical analog) the probability function is required to be that of the Born rule. On the interpretation advocated in this paper, this amounts to establishing that such an agent `cares about' the future branches in proportion to their amplitude-squared measures. (This means, among other things, that the amplitude-squared measure of a branch acts as a rational betting quotient, and (given an appropriate \emph{a priori} link between rational belief and rational action) satisfies Lewis' Principal Principle.)

 The proof is valid; whether we regard it as establishing its conclusion depends on whether we are antecedently more inclined to accept the premises than the conclusion itself. So: are we? As mentioned above, I have argued that the decision-theoretic axioms are perfectly acceptable, although the Everettian should justify them in the non-standard way sketched in section \ref{cf}. I regard Physicality as an obvious truth. Measurement Neutrality (MN), however, is the weak point of the proof: given the Everettian account of measurement, it is hard to see why one should accept MN as a primitive rationality constraint.

 I have suggested that we should accept the Born rule itself as something of a primitive; this suggestion is supported by the observation that there are no remotely plausible alternatives on offer. This approach has more modest ambitions than Deutsch's original project, but it, almost as well as Deutsch's and Wallace's approach, promises to solve the problems of probability in Everett. The Everett interpretation gives us coefficients that \emph{physically} govern interference effects; we should accept that interpretation, \emph{if} the link between rational action and theory confirmation conjectured in section \ref{link} obtains, because we \emph{also} take those coefficients to appropriately govern rational action. Whether or not that crucial condition is met, however, remains an open question.
 \end{section}

\section*{Acknowledgements}

For valuable discussions and/or correspondence, I am grateful to David Albert, Adam Elga, Barry Loewer, Oliver Pooley, Simon Saunders, an anonymous referee for Studies in the History and Philosophy of Modern Physics, and especially to Harvey Brown and David Wallace.

 \section*{References}

 \indent

 Albert, D. and Loewer, B. Interpreting the many-worlds interpretation. \emph{Synthese 77} (1988), 195-213.

 Barnum, H., Caves, C., Finkelstein, J., Fuchs, C., and Schack, R. (2000) Quantum probability from decision theory? \emph{Proceedings of the Royal Society of London A456} (2000), 1175-1182. Available online at http://www.arxiv.org/abs/quant-ph/9907024.

 Barrett, J. (1999) \emph{The quantum mechanics of minds and worlds.} Oxford: Oxford University Press.

 Bohm, D. (1952) A suggested interpretation of quantum theory in terms of `hidden' variables. \emph{Physical Review 85}, 166-93.

 Deutsch, D. (1985) Quantum Theory as a Universal Physical Theory. \emph{International Journal of Theoretical Physics 24(1)}, 1-41.

 Deutsch, D. (1999) Quantum Theory of Probability and Decisions. \emph{Proceedings of the Royal Society of London A455} (1999), 3129-37. Available online at http://www.arxiv.org/abs/quant-ph/9906015.

 DeWitt, B. (1970) Quantum Mechanics and Reality. \emph{Physics Today 23(9)} (1970), 30-35. Reprinted in DeWitt and Graham (1973).

 DeWitt, B. and Graham, N. (1973) \emph{The Many-Worlds Interpretation of Quantum Mechanics.} Princeton: Princeton University Press.

 Earman, J. (1986) \emph{A Primer on Determinism.} Reidel.

 Everett, H. (1957) `Relative state' formulation of quantum mechanics. \emph{Review of Modern Physics 29} (1957), 454-62. Reprinted in DeWitt and Graham (1973).

 Fishburn, P. (1981) Subjective Expected Utility: A Review Of Normative Theories. \emph{Theory and Decision 13} (1981), 139-99.

 Gell-Mann, M. and Hartle, J. (1990) Quantum mechanics in the light of quantum cosmology. In Zurek (ed.), \emph{Complexity, Entropy and the Physics of Information}. Redwood City, California: Addison-Wesley.

 Ghirardi, G. Rimini, A. and Weber, T. (1986) Unified dynamics for micro and macro systems. \emph{Physical Review D 34} (1986), 470.

 Graham, N. (1973) \emph{The Measurement of Relative Frequency.} In DeWitt and Graham (1973).

 Healey, R. (1984) How many worlds? \emph{Nous 18(4)} (1984), 591-616.

 Ismael, J. (2003) Chance and determinism. \emph{Philosophy of Science} 70, 776-790.

 Joyce, J. (1999) \emph{The foundations of causal decision theory.} Cambridge: Cambridge University Press.

 Kent, A. (1990) Against many-worlds interpretations. \emph{International Journal of Theoretical Physics A5} (1990), 1745-62. Available online at http://www.arxiv. \linebreak org/abs/gr-qc/9703089.

 Lewis, D. (1976) Survival and Identity. In Amelie O. Rorty (ed.), \emph{The Identities of Persons}, 17-40. University of California Press. Reprinted in Lewis, \emph{Philosophical Papers}, vol. I, Oxford University Press, 1983.

 Lewis, D. (1979) Attitudes De Dicto and De Se. \emph{The Philosophical Review 88} (1979), 513-43. Reprinted in Lewis, \emph{Philosophical Papers}, vol. I, Oxford University Press, 1983.

 Lewis, D. (2004) How many lives has Schrodinger's Cat? \emph{Australasian Journal of Philosophy 82(1)} (2004), 3-22.

 Lockwood, M. (1996) `Many Minds' Interpretations of Quantum Mechanics. \emph{British Journal for the Philosophy of Science 47} (1996), 159-188.

 Loewer, B. (1996) Comment on Lockwood. \emph{British Journal for the Philosophy of Science 47} (1996), 229-232.

 Papineau, D. (1996) Many Minds Are No Worse Than One. \emph{British Journal for the Philosophy of Science 47} (1996), 234-41.

 Parfit, D. (1984) \emph{Reasons and Persons.} Oxford: Oxford University Press.

 Pearle, P. (1989) Combining stochastic dynamical state-vector reduction with spontaneous localisation. \emph{Physical Review A 39(5)}, 2277-2289.

 Saunders, S. (1995) Time, Quantum Mechanics and Decoherence. \emph{Synthese 102} (1995), 235-266.

 Saunders, S. (1997) Naturalizing Metaphysics. \emph{The Monist 80(1)} (1997), 44-69.

 Saunders, S. (1998) Time, Quantum Mechanics and Probability. \emph{Synthese 114} (1998), 373-404. Available online at http://philsci-archive.pitt.edu/ \linebreak documents/disk0/00/00/ 04/65/index.html.

 Savage, L. (1972) \emph{The foundations of statistics.} New York: Dover.

 Teller, P. (1973) Conditionalization and observation. \emph{Synthese 26}:218-258.

 Vaidman, L. (1998) On schizophrenic experiences of the neutron. \emph{International Studies in the Philosophy of Science 12} (1998), 245-261. Available online at http://www.arxiv.org/abs/quant-ph/9609006.

 Vaidman, L. (2001) Many-worlds interpretation of quantum mechanics. \emph{Stanford encyclopaedia of philosophy.} Available online at http://plato.stanford.edu/ \linebreak entries/qm-manyworlds/.

 van Fraassen, B. (1984) Belief and the Will. \emph{Journal of Philosophy LXXXI} (1984), 235-256.

 von Neumann, J. and Morgenstern, O. (1947) \emph{Theory of Games and Economic Behaviour.} Princeton: Princeton University Press.

 Wallace, D. (2002a) Worlds in the Everett Interpretation. \emph{Studies in the History and Philosophy of Modern Physics 33(4)} (2002), 637-661. Available online at http://www.arxiv.org/abs/quant-ph/0103092.

 Wallace, D. (2002b) Quantum Probability and Decision Theory, Revisited. Available online at http://www.arxiv.org/abs/quant-ph/0211104.

 Wallace, D. (2003a) Everett and Structure. \emph{Studies in the History and Philosophy of Modern Physics 34(1)} (2003), 87-105. Available online at http://www. \linebreak arxiv.org/abs/quant-ph/0107144.

 Wallace, D. (2003b) Quantum Probability and Decision Theory, Revisited. Forthcoming in \emph{Studies in the History and Philosophy of Modern Physics}. Available online at http://www.arxiv.org/abs/quant-ph/0303050, under the title `Everettian Rationality: defending Deutsch's approach to probability in the Everett interpretation.' (This paper is a shorter version of Wallace (2002b).)

 Wallace, D. (2003c) Quantum probability from subjective likelihood: improving on Deutsch's proof of the probability rule. Available online at http://www.arxiv.org/abs/quant-ph/0312157.

 Wilkes, K. (1988) \emph{Real People.} Oxford: Clarendon Press.

 Zurek, W. H. (1991) Decoherence and the transition from quantum to classical. \emph{Physics Today 44}, 36-44.

 Zurek, W. H. (1993) Preferred states, predictability, classicality and the environment-induced decoherence. \emph{Progress of Theoretical Physics 89(2)}, 281-312.

 Zurek, W.H. (1998) Decoherence, Einselection and the Existential Interpretation (The Rough Guide). \emph{Philosophical Transactions of the Royal Society of London A356}, 1793-1820.

 Zurek, W. H. (2003) Decoherence, einselection and the quantum origins of the classical. \emph{Reviews of Modern Physics 75(3)} (July 2003), 715-775. Available online at http://www.arxiv.org/abs/quant-ph/0105127.

 \end{document}